\documentclass[a4paper,10pt]{article}
\pdfoutput=1 
\usepackage{jcappub} 
\usepackage{mathrsfs} 
\usepackage{amsmath,mathtools}
\newcommand{\hp}{\hphantom0}
\usepackage{graphicx}
\usepackage{subcaption}

\usepackage{dcolumn}
\usepackage{bm}
\def\be{\begin{equation}}
\def\ee{\end{equation}}
\def\bea{\begin{eqnarray}}
\def\eea{\end{eqnarray}}

\usepackage{slashed}
\usepackage{gensymb}
\def\lsim{\raise0.3ex\hbox{$\;<$\kern-0.75em\raise-1.1ex\hbox{$\sim\;$}}}
\def\gsim{\raise0.3ex\hbox{$\;>$\kern-0.75em\raise-1.1ex\hbox{$\sim\;$}}}

\usepackage{pstricks}
\usepackage{epsfig}

\title{\boldmath Gauge-Field-Induced Torsion and Cosmic Inflation}
\author[a]{A.~Kasem}
\author[b]{and S.~Khalil}
\affiliation[a, b]{\small Center for Fundamental Physics, Zewail City of Science and Technology, 6 October City, Giza 12578, Egypt.}

\emailAdd{amkassem@zewailcity.edu.eg}
\emailAdd{skhalil@zewailcity.edu.eg}
\abstract{
Inflation in the framework of Einstein-Cartan  theory is revisited. Einstein-Cartan theory is a natural extension of the General Relativity, with non-vanishing torsion. The connection on Riemann-Cartan spacetime is only compatible with the cosmological principal for a particular form of torsion. We show this form to also be compatible with gauge invariance principle for a non-Abelian and Abelian gauge fields under a certain deviced minimal coupling procedure. We adopt an Abelian gauge field in the form of ``cosmic triad''. The dynamical field equations are obtained and shown to sustain cosmic inflation with a large number of e-folds. We emphasize that at the end of inflation, torsion vanishes and the theory of Einstein-Cartan reduces to the General Relativity with the usual FRW geometry.
}
\begin{document}
\maketitle
\flushbottom
\section{Introduction}

From the outset of the general theory of relativity (GR), it has been playing a major role as a framework for our thinking about the universe. The theory has witnessed many success stories throughout the past century that have been crowned by what is known as the standard model of cosmology or $\Lambda CDM$ model \cite{Dicke:1965zz, PhysRevD.23.347, Peebles:1982ff, Riess:1998cb}. The model describes a homogeneous, isotropic, and spatially-(almost)-flat universe started at a very dense state ``big bang'', undergone a period of exponential expansion and has been expanding in an accelerated manner ever since. Despite what the name suggests, the model is plagued with many theoretical issues. It is only standard as a best fit model to the observational evidence we have.

Problems with $\Lambda CDM$ model are de facto inherited from GR which is still struggling to find a foothold in the general scheme of particle physics. Success of GR in describing large scale physics should not tempt us to overlook other alternatives. Having said that, it is more likely the alternative being an extension to GR so one is guaranteed not to stray from observational tests.

We argue that a good, well-motivated candidate would be Einstein-Cartan theory. The theory differs by admitted a torsion tensor in space, an object conspicuous by its absence in GR. Once one starts by building a local gauge theory of Poincar\'{e} group, Einstein-Cartan falls into place as the other forces in particle physics with no a priori reason for having a vanishing torsion.

Torsion can be a rather de trop object. The antisymmetic part of the connection explicitly violates gauge invariance when introducing a gauge field into curved background. In the cosmological context, several semi-classical models have been regrading torsion as a spinning fluid \cite{PhysRevD.35.1181, Gasperini:1986mv, Tsoubelis:1979zz, Ray:1982qr, Bradas:1987mk, Poplawski:2010kb}. In some cases they required an unreasonable amount of fine tuning or an extremely large amount of spinning particles in the early universe. But they were mainly not successful because torsion violated homogeneity and isotropy of space and inhomogeneous cosmology has not yet borne fruit on the observational level.

In this paper, we aim to introduce a scenario of torsion sourced by a gauge field. We show that a minimal setup motivated by respecting FRW symmetries and gauge invariance sustains a period of exponential expansion. We start in section \ref{Einstein-Cartan} by looking at the development of Einstein-Cartan theory as gauge theory of gravity. In section \ref{Source of torsion}, we see that only a certain form of torsion is compatible with FRW symmetries and it can be sourced by a gauge field provided that a new minimal coupling prescription is adopted. In section \ref{Model with cosmic triad}, we setup the model and compute the field equations. In section \ref{Solutions}, we show that the field equations can sustain a period of early cosmic inflation with a large number of e-folds. Our conclusions and final remarks are given in section \ref{conclusion}.

Appendixes \ref{maximally-symmetric torsion} and \ref{AField Strength} provide calculations for the form of torsion compatible with FRW symmetries and gauge invariance of a nonabelian gauge field. In appendixes \ref{AEnergyMomentum} and \ref{ARIcci and E tensors} we provide the explicit form of the energy momentum, Ricci, and Einstein tensors that we used to compute the field equations.

\section{Development of Einstein-Cartan Theory}\label{Einstein-Cartan}
The concept of an antisymmetric affine connection was first introduced by \'{E}lie Cartan in 1922, later development of a gravitational theory allowing torsion did not ensue much until the sixties. This in part was due to the consensus on Einstein GR which passed all macrophysical experimental tests. Later development of particle physics made it clear that a direct measurement of gravitational effects between elementary particles is beyond our technological capabilities. However, it gave us a general scheme to govern all forces of nature. Gravity can only fit into the general scheme of things if it can be written as a gauge theory.

If one thinks of a global symmetry for gravity, the group of Lorentz transformations is the obvious candidate. Promoting this symmetry into local one was an endeavor first started by Utiyama \cite{Utiyama:1956sy}, and later developed by Kibble \cite{Kibble:1961ba} and Sciama \cite{Sciama1962OnTA, RevModPhys.36.463}. The theory is better known as Einstein-Cartan-Sciama-Kibblle or $U_4$ for short. $U_4$ is indeed the local gauge theory of the Poincar\'{e} group.\footnote{For details and accurate historical development, readers are refereed to \cite{RevModPhys.48.393, Blagojevic:2002du}.} It features a connection with an antisymmetric part known as torsion; $S^i_{\hp jk}\equiv \Gamma^i_{\hp [jk]}=\frac{1}{2}(\Gamma^i_{\hp jk}-\Gamma^i_{\hp kj})$. With the help of the metric postulate, one can write the connection in terms of the metric and torsion as follows
\begin{equation}
\Gamma^i_{\hp jk}=\tilde{\Gamma}^i_{\hp jk}+K^i_{\hp jk}.
\end{equation}
Where $\tilde{\Gamma}^i_{\hp jk}$ being the usual Levi-Civita Connection and the contortion tensor $K^i_{\hp jk}$ is defined as $K^i_{\hp jk}\equiv S^i_{\hp jk}+S^{\hp i}_{k \hp j}-S^{\hp\hp i}_{jk}$. As the case with GR, one can still construct an invariant scalar quantity using the Riemann curvature tensor so the simplest Lagrangian would be $\mathscr{L}=\sqrt{-g}(\frac{1}{2}R(\Gamma)+\mathscr{L}_M)$. By a straightforward but rather lengthy calculations, one finds the field equations by varying the action with respect to the independent variables ($10$ represented by the metric and $24$ represented by the torsion) resulting in the field equations;
\begin{align}
G^{\mu\nu}&=\mathscr{T}^{\mu\nu}\label{Einstein Field eq.}\\
T^{\alpha\beta\gamma}&=\mathscr{S}^{\alpha\beta\gamma},\label{torsion algeb. field eq}
\end{align}
where $\mathscr{T}^{\mu\nu}$ ``Energy-momentum tensor'', $T^{\alpha\beta\gamma}$ ``modified torsion tensor'', and $\mathscr{S}^{\alpha\beta\gamma}$ ``spin energy potential'' are defined as follows;
\begin{equation}\label{Canonical EM}
\mathscr{T}^\mu_\nu\equiv q_{,\nu}\frac{\partial\mathscr{L}_M}{\partial q_{,\mu}}-\delta^\mu_\nu\mathscr{L}_M,
\end{equation}
\begin{equation}
T_{\hp\alpha\beta}^{\gamma}\equiv S_{\hp\alpha\beta}^{\gamma}+\delta^\gamma_\alpha S_{\hp\beta \lambda}^{\lambda}-\delta^\gamma_\beta S_{\hp\alpha \lambda}^{\lambda},
\end{equation}
\begin{equation}\label{torsion spin tensor}
\mathscr{S}^{\alpha\beta\gamma}\equiv\frac{1}{2\sqrt{-g}}\big(\frac{\delta\mathscr{L}_M}{\delta S_{\alpha\beta\gamma}}-\frac{\delta\mathscr{L}_M}{\delta S_{\alpha\gamma\beta}}\big).
\end{equation}
Unless one acquiesces to setting the torsion to zero, it should be regarded as fundamental as the metric itself. It is important to note that GR cannot be written as a gauge theory \cite{Aldrovandi:2013wha}. One can reach the field equations of GR with the local theory of translations with torsion as a gauge field strength. However, this formulation is only equivalent at the level of field equations as TEGR action involves a torsion scalar rather than a curvature scalar and the theory is formulated with a Weitzenb\"{o}ck connection rather than a Levi-Civita connection.

Few remarks on the field equations of $U_4$ are in order. First we note from the second field equation (\ref{torsion algeb. field eq}) is an algebraic equation so it can be fed into Eq.(\ref{Einstein Field eq.}) so one recovers Einstein field equations with an ``effective energy momentum tensor''.
Secondly, the canonical Energy-momentum (\ref{Canonical EM}) is not symmetric for a spinor field, i.e. for Dirac field it is totally antisymmetric ${T^\mu_\nu}_{\text{can}}=\frac{i}{2}\bar{\Psi}\gamma^\mu\partial_\nu \Psi-\frac{i}{2}\Psi\gamma^\mu\partial_\nu \bar{\Psi}$.\footnote{It is only upon subtraction from the divergence of the spin tensor, one gets a symmetric tensor equivalent to the gravitational energy momentum; namely the Belinfante tensor.} This may give us some account as of why torsion is linked to spin in terminology. A bosonic field does not feel rotation in space so it cannot source a torsion tensor. The type of possible matter fields feeling torsion is a topic we address in the next section.

\section{Source of torsion}\label{Source of torsion}

Early cosmological treatments of $U_4$ adopted a semiclassical model of a spinning dust known as Weyssenhoff fluid \cite{Weyssenhoff:1947iua}. It features a singularity-free solution \cite{Trautman:1973wy, Kopczynski:1972fhu, osti_4165396, Gasperini:1998eb}, and has been studied in the context of inflation \cite{Gasperini:1986mv, PhysRevD.35.1181, Poplawski:2010kb}. A number of these models have underplayed the incompatibility of this fluid with the symmetries of FRW universe \cite{TSAMPARLIS197927, Boehmer:2006gd}. To get more insight we can look at the torsion covariantly split into three irreducible parts \cite{Capozziello:2001mq};
\begin{align}
S_{\hp ab}^{c}&={}^{T}S_{\hp ab}^{c}+{}^{A}S_{\hp ab}^{c}+{}^{V}S_{\hp ab}^{c},
\end{align}
where ${}^{V}S_{\hp ab}^{c}$, ${}^{A}S_{\hp ab}^{c}$, and ${}^{T}S_{\hp ab}^{c}$ are vector, axial vector and tensorial components; respectively. They are defined as follows
\begin{align}
{}^{V}S_{\hp ab}^{c}&\equiv\frac{1}{3}(S_a\delta_b^c-S_b\delta_a^c),\label{Vtorsion}\\
{}^{A}S_{\hp ab}^{c}&\equiv g^{cd}S_{[abd]},\\
{}^{T}S_{\hp ab}^{c}&\equiv S_{\hp ab}^{c}-{}^{A}S_{\hp ab}^{c}-{}^{V}S_{\hp ab}^{c},
\end{align}
where the one-indexed $S_a$ is defined as $S_a\equiv S_{\hp ab}^{b}$. We wish to only keep the parts in $S^c_{\hp ab}$ that make it form-invariant on a spatially maxiamally-symmetric space. To this end we require the vanishing of the Lie derivative with respect to the 6 Killing vectors of that symmetry; $L_\xi S^a_{ab}=0$. In Appendix \ref{maximally-symmetric torsion}, we present a succinct proof following the work of \cite{TSAMPARLIS197927}. The result is a vanishing tensorial part ${}^{T}S_{\hp ab}^{c}=0$ and a particular form for the vector and axialvector parts, namely
\begin{equation}\label{VandA torsion}
  \begin{split}
S^1_{\hp 01}=S^2_{\hp 02}=S^3_{\hp 03}=F(t), \\
S_{123}=S_{[123]}=f(t),
  \end{split}
\end{equation}
where $F(t)$ and $f(t)$ are arbitrary functions of time. It is evident that this form is incompatible with the spinning fluid description which has a non-vanishing traceless part. The cosmological implications of torsion in the form of (\ref{VandA torsion}) have been investigated in various works \cite{PhysRevD.24.1451, Kranas:2018jdc, Pereira:2019yhu, Bose:2020mdm, Goenner_1984}. 
As this type of models undermine a spinning fluid description, it gives up one of its significant results; a universe free of cosmic singularity. Moreover, the functions $F(t)$, $f(t)$ are introduced in an ad hoc manner with no available physical interpretation to its source. Therefore,  one would be reluctant to take up this pursuit any further, however we show that the vector torsion part can be a grist to the mill of cosmology if it couples to a gauge field.

The possibility of sourcing torsion with a gauge field is often dismissed as torsion is known to violate gauge invariance, i.e. if one tries the natural generalization of the field strength of gauge field (Abelian or non-Abelian) to Einstein-Cartan manifold through replacing the partial derivative by a space-time derivative: $\partial_\mu A_\nu\rightarrow \partial_\mu A_\nu-\Gamma^\lambda_{\hp\mu\nu}A_\lambda$, one finds 
\begin{align}\label{Field strength nabla}
F^i_{\mu\nu}&\equiv\nabla_\mu A^i_\nu-\nabla_\nu A^i_\mu+g C^i_{\hp jk}A^j_\mu A^k_\nu\nonumber\\&=\partial_\mu A^i_\nu-\partial_\nu A^i_\mu-S^\lambda_{\hp\mu\nu}A^i_\lambda+g C^i_{\hp jk}A^j_\mu A^k_\nu.
\end{align}
Thus, gauge invariance is found to be manifestly violated due to the antisymmetric part of the connection. One way to avoid this violation is adopting the flat-space expression for $F_{\mu\nu}$. However, this defeats our purpose in introducing this field in the first place.

A more satisfactory approach have been introduced by Hojman, et al. \cite{PhysRevD.17.3141}, where a new minimal coupling procedure has been devised to allow for a nonvanishing torsion in the definition of an Abelian gauge field strength. Then, it  was shorty generalized for a non-abelian field as well \cite{Mukku:1978mz}. In this procedure, one defines the gauge transformation to allow a set of space-time functions $b_\mu^{\hp\alpha}$ into the the definition of a covariant derivative such that; $D_\mu=\nabla_\mu-igb_\mu^{\hp\alpha}A_\alpha$. This affects the transformation of the gauge field in the following way 
\begin{align}
{A^i_\mu}^\prime&=A^i_\mu-g^{-1}c_\mu^{\hp\nu}\theta^i_{;\nu}+C^i_{\hp jk}\theta^jA^k_\mu,
\end{align}
where $c_\mu^{\hp\nu}b_\nu^{\hp\alpha}\equiv\delta^\alpha_\mu$. Now the following modified field strength can be made invariant (see Appendix \ref{AField Strength} for detailed calculations):
\begin{equation}\label{nonab F}
F^i_{\mu\nu}\equiv \partial_\mu A^i_\nu-\partial_\nu A^i_\mu-S^\beta_{\hp\mu\nu}A^i_\beta+ge^{-f} C^i_{\hp jk}A^j_\mu A^k_\nu,
\end{equation}
under the constraints of $c_\mu^{\hp\nu}=e^{f(X)}\delta_\mu^{\hp\nu}$ and $S^\sigma_{\hp\mu\nu}\equiv\delta_\nu^{\hp\sigma}f_{,\mu}-\delta_\mu^{\hp\sigma}f_{,\nu}$, which coincides with the expression of vector torsion Eq.(\ref{Vtorsion}). It also conforms with the FRW-compatible torsion when one assumes $f$ as a function of time $f(X)\equiv f(t)$.
One can think of this procedure as giving a nonconstant gauge coupling $g(t)=ge^{-f(t)}$. So in the limit of vanishing torsion, the usual minimal coupling procedure is restored with $c_\mu^{\hp\nu}=\delta_\mu^{\hp\nu}$ and $g(t)\equiv g$. Our discussion extends to three Abelian fields $A_\mu^i$ which can be thought of as a special case of $SU(2)$ nonabelian gauge theory with vanishing structure constants $C^i_{\hp jk}\rightarrow0$. For the purpose of constructing a minimal setup, we stick with the Abelian version in the following analysis.

\section{Model with cosmic triad}\label{Model with cosmic triad}
Having established an FRW-compatible form of torsion, we would like to couple it with gauge gauge fields respecting the same symmetries. We adopt an ansatz of three equal and mutually-orthogonal vectors known as ``cosmic triad'' was proposed by \cite{ArmendarizPicon:2004pm};
\begin{equation}
A^a_i=\phi(t)\cdot a(t)\delta^a_i ,
\end{equation}
where $a(t)$ is the FRW scale factor. The cosmological implication of this choice has been investigated in many different contexts \cite{Wei:2006gv, Bamba:2008xa, Elizalde:2010xq, Maleknejad:2011sq, Elizalde:2012yk, Mehrabi:2015lfa, Rinaldi:2015iza, Rodriguez:2017wkg, Alvarez:2019ues}. Here we take the simplest form with three identical copies of an Abeilan field coupled with torsion as per the prescription of the previous section;
\begin{equation}\label{F0i}
F^a_{0i}=(a\dot{\phi}+\dot{a}\phi)\delta^a_i-\phi a\dot{f}\delta^a_i
\end{equation}
The standard Lagrangian $\mathscr{L}=\sqrt{-g}\big[\frac{1}{2}(R-2\Lambda)-\frac{1}{4}F^a_{\mu\nu}F_a^{\mu\nu}+\mathscr{L}_M\big]$ is constructed assuming FRW background and the aformentioned choice for torsion and fields. We assume a flat space, a vanishing cosmological constant and a universe dominated by the triad field. Also we relabel the torsion function $\dot{f}\rightarrow \chi(t)$. Now the above Lagrangian simplifies into;
\be
\mathscr{L} = 3a\dot{a}^2-12\chi^2a^3+\frac{3a}{2}(\dot{a}\phi+a\dot{\phi}-\phi a \chi)^2
\ee

We calculate the field equations (\ref{Einstein Field eq.}, \ref{torsion algeb. field eq}), using the energy-momentum tensor and the Einstein tensor as given in appendix \ref{ARIcci and E tensors}, and \ref{AEnergyMomentum}. A straightforward but rather onerous calculations gives the following Freidman-like equations and an extra equation for the torsion field:
\bea
\big(\frac{\dot{a}}{a}\big)^2&=&\frac{1}{3}\rho_A-4\chi\frac{\dot{a}}{a}-4\chi^2, \label{Friedmann 1}\\
\frac{\ddot{a}}{a}&=&-\frac{1}{6}(\rho_A+3P_A)-2\chi\frac{\dot{a}}{a}-2\dot{\chi}, \label{Friedmann 2}\\
\ddot{\phi}&=&-(3H+\chi) \dot{\phi}+(2\chi^2+\dot{\chi}-2H^2-\dot{H})\phi,
\eea
with 
\be 
\chi=\frac{1}{2a^3}\frac{\delta\mathscr{L}_M}{\delta\chi}\Rightarrow \chi=\frac{3\phi^2}{3\phi^2-2}(\frac{\dot{\phi}}{\phi}+H).
\label{torsion from phi}
\ee
As we expect, the torsion equation (\ref{torsion from phi}) is algebraic, so one can use it to substitute for the nondynamical field. By introducing the Hubble parameter $H(t)\equiv\frac{\dot{a}}{a}$ and explicitly substitute for $\rho_A$ and $P_A$, defined in Appendix \ref{AEnergyMomentum}, one finds that the above equations can be written as 
\bea 
H^2&=&2\frac{H\phi+\dot{\phi}}{(2-3\phi^2)^2}\big( (1-18\phi^2)\dot{\phi}+(13-36\phi^2)H\phi \big), \label{Friedmann 1'}\\
\dot{H}+H^2&=&\frac{2}{(2-3\phi^2)^2}\big( (5+9\phi^2)\dot{\phi}^2-9(H^2+\dot{H})\phi^4+(5H^2+6\dot{H})\phi^2-9(H\dot{\phi}+\ddot{\phi})\phi^3\nonumber\\
&+& 2(8H\dot{\phi}+3\ddot{\phi})\phi \big),\label{Friedmann 2'}\\
\ddot{\phi}&=&\frac{H}{(2-3\phi^2)^2}\Big( 3(2-5\phi^2)\dot{\phi}+4(1-3\phi^2)H\phi \Big)-\phi\dot{H}. \label{phi eq}
\eea

\section{Inflationary Scenario}\label{Solutions}
To understand the cosmic evolution let us start with Eq.(\ref{Friedmann 1'}), which is a quadratic equation of the Hubble parameter, one can solve to get $H(\phi,\dot{\phi})$
\begin{equation}\label{HSolution}
H=\frac{\pm 2\sqrt{2}+(14-3(18\phi\pm\sqrt{2})\phi)\phi}{4-38\phi^2+81\phi^4}\dot{\phi}.
\end{equation}
Substituting this solution as well as $\ddot{\phi}$ from Eq.(\ref{phi eq}) into Eq.(\ref{Friedmann 2'}), one finds $\dot{H}(\phi,\dot{\phi})$
\begin{equation}\label{Hd}
\dot{H}=\frac{\mp 2\dot{\phi}^2}{(4-38\phi^2+81\phi^4)^2}\Big(\mp 16+\big(32\sqrt{2}+(\pm 412+3(-52\sqrt{2}+3(\mp 254+9(2\sqrt{2}\pm 45\phi)\phi)\phi)\phi)\phi\big)\phi  \Big)
\end{equation}

\begin{figure}[t!]
 \centering
\includegraphics[width=0.5\textwidth]{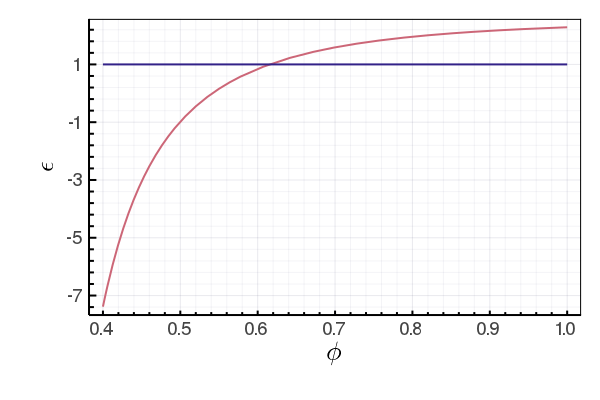}
\caption{The inflationary parameter $\epsilon$ as function of the field $\phi$}
\label{analytical par.}
\end{figure}

Now we are set to find the inflationary parameter $\epsilon\equiv\frac{\dot{H}}{H^2}$, which we use to identify the inflationary period $\ddot{a}>0$ as $\frac{\ddot{a}}{a}=H^2(1-\epsilon)$, so exponential growth is going in the region of $\epsilon<1$. From the above expressions of $H$ and $\dot{H}$, one obtains the following relation of $\epsilon$ in terms of $\phi$ scalar field. 
\begin{equation}\label{epsilon}
\epsilon\equiv-\frac{\dot{H}}{H^2} =-\frac{8\mp 24\sqrt{2}\phi-90\phi^2}{(\sqrt{2}\pm 6\phi^2)^2}.
\end{equation}
We stress that we only use this parameter to define the inflationary period with no requirement on epsilon as to being positive or infinitesimal. For reasons that should be obvious shortly, we find only the second solution can sustain a period of inflation large enough to generate a desirable number of e-fold so we stick to it in the following analysis. In Fig. \ref{analytical par.}, we plot inflationary parameter $\epsilon$ versus the scalar field $\phi$. As can be seen from this figure, the inflationary period ($\epsilon<1$) is going on as long as the field is less than a certain value $\phi <\frac{1}{9}(\sqrt{2}+\sqrt{17})\approx0.62$. It is clear that for a large field inflation, i.e. $\phi > 1$, inflation starts when $\phi$ is rolling down to vales less than $0.62$, while for a small field, i.e. $\phi$ starts from values close to zero and rolls up, the inflation ends when $\epsilon$ approaches one at $\phi \simeq 0.62$. 

\begin{figure}[t!]
\includegraphics[width=0.5\textwidth]{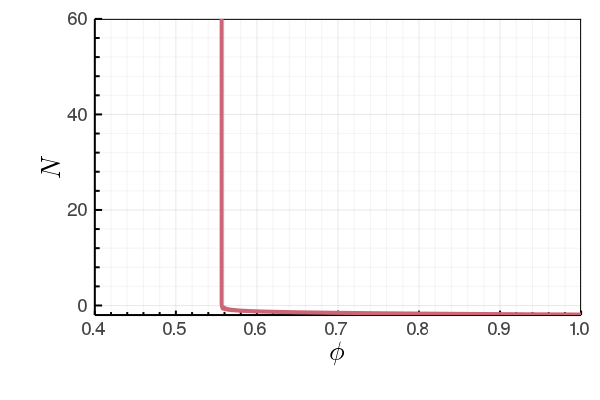} ~~ \includegraphics[width=0.5\textwidth]{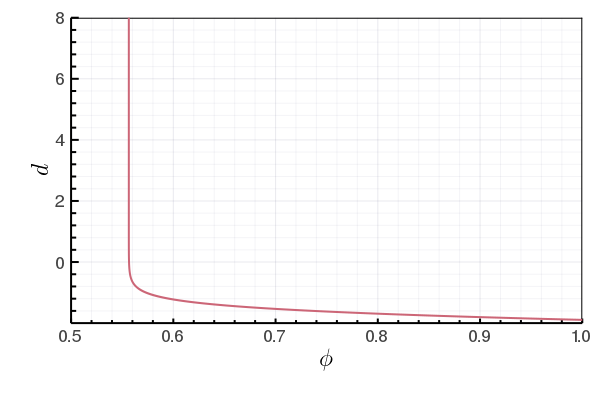}
 \caption{The integral $N$ and the drag function $d$ versus  the field $\phi$}
\label{analytical par2}
\end{figure}

Now, we consider the number of e-folds $N$ of inflation that are needed to solve the horizon problem. In general $N$ is defined as $N \equiv \int_{t_i}^{t_f}  H dt$. In our model, one can show that it is given by 
\begin{equation}
N=\frac{1}{111}\Big((2\sqrt{37}-37)\log(\sqrt{2}+\sqrt{74}-18\phi)-(2\sqrt{37}+37)\log(-\sqrt{2}+\sqrt{74}+18\phi)\Big)\Biggr|_{\phi_i}^{\phi_f}
\end{equation}
In Fig. \ref{analytical par2}, we  show the integral of e-folds which has a notable pole at $\phi^*=\frac{1}{18}(\sqrt{2}+\sqrt{74})\approx 0.556$, and the required value $N\simeq 60$ can be obtained when the field $\phi$ approach this point. In this case, we have a new scenario of inflation, where  the field $\phi$ is rolling down from a relatively large value with an initial velocity, an inflationary period starts at $\phi_i$ (the point for $\epsilon=1$) and as the field approaches the pole at $\phi^*$ the number e-folds blows up, as shown in Fig. \ref{analytical par2}, and inflation abruptly ends. 

The abrupt end of the inflation can be understood by considering the evolution of the field $\phi$, by substituting Eq.(\ref{HSolution}) and Eq.(\ref{Hd}) into Eq.(\ref{phi eq}), one gets the following equation for the evolution of the scalar field $\phi$
\begin{equation}\label{phi eq'}
\ddot{\phi}=\pm\frac{-6\sqrt{2}+2(\mp18+(\sqrt{2}\pm72\phi)\phi)\phi}{4-38\phi^2+81\phi^4}\dot{\phi^2}.
\end{equation}
Thus, as the field rolls down, it suffers a non-constant drag of $d(\phi)=\frac{6\sqrt{2}-2(18+(\sqrt{2}-72\phi)\phi)\phi}{4-38\phi^2+81\phi^4}$. As one notices from Fig. \ref{analytical par2}, the drag approaches an infinite value at the point $\phi^*$ creating a virtual infinite barrier at this point forcing the field to stop. As the field stops, i.e.  $\dot{\phi}\approx 0$, inflation ends as $H\sim \dot{\phi}\approx 0$. How close the field gets to $\phi^*$ determines how much N-folds accumulated and depends on the initial conditions for the field and its velocity.

It worth noting that by end of inflation, torsion will vanish (as $\dot \phi =0$), and thus the theory of Einstein-Cartan reduces to the GR with the usual FRW geometry. Therefore, in this scenario the torsion begins with large values at very early times of the universe that give rise inflation. Then, it diminishes at the end of inflation, in conformity with our late-time universe.  

\section{Conclusions}\label{conclusion}
In this paper we have analyzed a novel scenario for cosmic inflation induced by a gauge field in the framework of Einstein-Cartan theory. We show that the vector torsion can play a crucial role in the early cosmology if it couples to a gauge field. We show a specific form of the torsion, in terms of one scalar function, complies with both the cosmological principal (spatial homogeneity and isotropy) and gauge invariance principle. We consider a simple example of three equal and mutually orthogonal abelian gauge fields, in the form of ``cosmic triad".  We calculate the corresponding field equations and show that torsion is determined, through an algebraic equation,  in terms of the scalar field of the cosmic triad. By solving the dynamical field equations, we found that a period of inflation can be sustained.

The field evolution corresponds to large field (chaotic) type of inflation as the large number of e-folds accumulated near the end of inflation depends on the initial conditions. At the end of inflation, the field has a constant background value while torsion vanishes and the theory reduces to the General Relativity with the usual FRW geometry. Finally, we would like to point out prospects of studying this model in light of cosmological perturbation theory and possible reheating mechanisms.

\newpage
\acknowledgments The authors would like to thank A. Awad, A. El-Zant, A. Golovnev, E. Lashin, and G. Nashed for the useful discussions.\\

\appendix 
\hspace{-0.85cm} {\Large \bf Appendix}
\section{Spatially Maximally form-invariant 3-rank
antisymmetric tensor}\label{maximally-symmetric torsion}

We assume that we have a maximally symmetric subspace. It is
always possible to find a coordinate transformation for which the metric of the whole space take the following form \cite{Weinberg:1972kfs}

\begin{equation}\label{ds Maximal metric}
ds^2 = g_{\mu\nu}dx^\mu dx^\nu = g_{ab}(V)dv^a dv^b + f(V) \tilde{g}_{ij}(U)du^i du^j
\end{equation}

Where $\tilde{g}_{ij}(U)$ being the metric of an $M$ dimensional maximally symmetric space. The indices $a,~b;\cdots$ label the invariant coordinates $V$ and run over $N-M$ while the indices $i,~j,\cdots$ label the infinitesimally
transformed coordinates $U$ and run over the $M$ labels.

A tensor $T_{\mu\nu\cdots}(X)$ is called maximally form-invariant if it has a vanishing Lie derivative with respect to the Killing vectors $\xi^\lambda(X)$;

\begin{equation}
L_\xi T_{\mu\nu\cdots}\equiv \xi^\lambda\frac{\partial}{\partial x^\lambda}T_{\mu\nu\cdots}+\frac{\partial\xi^\rho}{\partial x^\mu}T_{\rho\nu\cdots}+\frac{\partial\xi^\sigma}{\partial x^\nu}T_{\mu\sigma\cdots}+\cdots=0
\end{equation}

We apply this condition on an antisymmetric tensor $S^\mu_{\hp\nu\rho}=-S^\mu_{\hp\rho\nu}$ on the space of \ref{ds Maximal metric};

\begin{align}
L_\xi S^i_{\hp jk}=0\label{1st Killing}\\
L_\xi S^i_{\hp ja}=0\label{2nd Killing}\\
L_\xi S^i_{\hp ab}=0\label{3rd Killing}\\
L_\xi S^a_{\hp bc}=0\label{4th Killing}\\
L_\xi S^a_{\hp bi}=0\label{5th Killing}\\
L_\xi S^a_{\hp ij}=0\label{6th Killing}\\
\end{align}

The condition \ref{1st Killing} leads to

\begin{equation}
S^l_{\hp jk}\delta^m_i+S^{\hp l}_{i\hp k}\delta^m_j+S^{\hp\hp l}_{ij}\delta^m_k=S^m_{\hp jk}\delta^l_i+S^{\hp m}_{i\hp k}\delta^l_j+S^{\hp\hp m}_{ij}\delta^l_k
\end{equation}

Contracting $m$ and $i$, one gets

\begin{equation}
mS^l_{jk}+S^{\hp l}_{j\hp k}+S^{\hp\hp l}_{kj}=S^l_{\hp jk}+S^{\hp i}_{i\hp k}\delta^l_j+S^{\hp\hp i}_{ij}\delta^l_k
\end{equation}

Further contraction for $l$ and $j$ gives $T^i_{\hp ik}= 0$, substituting this back in the previous equation gives;

\begin{equation}
(m-1)S^l_{\hp jk}+S^{\hp l}_{j\hp k}+S^{\hp\hp l}_{kj}=0
\end{equation}

The solution for which is simply in the form of Levi-Civita tensor times an arbitrary function of time

\begin{equation}
S_{ijk}=\left\{
\begin{array}{ll}
      F(t)\epsilon_{ijk} & m=3 \\
      0 & m\neq 3 \\
\end{array} 
\right.
\end{equation}

The condition \ref{2nd Killing} implies

\begin{equation}
S^\alpha_{\hp j0}\delta^\beta_i+S^{\hp\alpha}_{i\hp 0}\delta^\beta_j+S^{\hp\hp\alpha}_{ij}\delta^\beta_0=S^\beta_{\hp j0}\delta^\alpha_i+S^{\hp\beta}_{i\hp 0}\delta^\alpha_j+S^{\hp\hp\beta}_{ij}\delta^\alpha_0
\end{equation}

Solution is zero unless $\alpha= j$ and $\beta=i$ for which;

\begin{equation}
S^i_{\hp i0}=S^j_{\hp j0}=f(t)~\forall~i,~j
\end{equation}

For \ref{3rd Killing} and \ref{4th Killing}, its trivial to show that $T^i_{\hp 00} = T^0_{\hp 00}= 0$. The fifth condition \ref{5th Killing} gives;

\begin{equation}
S^\alpha_{\hp 0i}\delta^\beta_0+S^{\hp\alpha}_{0\hp i}\delta^\beta_0+S^{\hp\hp\alpha}_{00}\delta^\beta_i=S^\beta_{\hp 0i}\delta^\alpha_0+S^{\hp\beta}_{0\hp i}\delta^\alpha_0+S^{\hp\hp\beta}_{00}\delta^\alpha_i\Rightarrow S^{\hp\hp k}_{00}\delta^j_i=S^{\hp\hp j}_{00}\delta^k_i\Rightarrow S_{00i}=0
\end{equation}

And the last condition \ref{6th Killing} gives;

\begin{equation}
S^\alpha_{\hp ij}\delta^\beta_0+S^{\hp\alpha}_{0\hp j}\delta^\beta_i+S^{\hp\hp\alpha}_{0i}\delta^\beta_j=S^\beta_{\hp ij}\delta^\alpha_0+S^{\hp\beta}_{0\hp j}\delta^\alpha_i+S^{\hp\hp\beta}_{0i}\delta^\alpha_j\Rightarrow S^{\hp k}_{0\hp (j}\delta^l_{i)}=S^{\hp l}_{0\hp (j}\delta^k_{i)}\Rightarrow S_{0ij}=0
\end{equation}

And this proves Eq.(\ref{VandA torsion}). The enthusiastic reader is encouraged to look at the discussion of maximally symmetric spaces in \cite{Weinberg:1972kfs, Carroll:1997ar}.

\section{Field Strength}\label{AField Strength}

The commutation of two covariant derivatives can be found as;

\begin{align}
[D_\mu,D_\nu]\psi&=[\nabla_\mu,\nabla_\nu]\psi-ig\nabla_\mu(b_\nu^{\hp\alpha}A_\alpha)\psi+ig\nabla_\nu(b_\mu^{\hp\alpha}A_\alpha)\psi-igb_\nu^{\hp\alpha}A_\alpha\psi_{;\mu}-igb_\mu^{\hp\alpha}A_\alpha\psi_{;\nu}\nonumber\\&-g^2b_\mu^{\alpha}b_\nu^{\beta}A_\alpha A_\beta\psi+igb_\mu^{\hp\alpha}A_\alpha\psi_{;\nu}+igb_\nu^{\hp\alpha}A_\alpha\psi_{;\mu}+g^2b_\nu^{\alpha}b_\mu^{\beta}A_\alpha A_\beta\psi\nonumber\\
&=-ig(\partial_\mu (b_\nu^{\hp\alpha} A^i_\alpha)-\partial_\nu (b_\mu^{\hp\alpha}A^i_\alpha)-S^\beta_{\hp\mu\nu}b_\beta^{\hp\alpha}A^i_\alpha+gC^i_{\hp jk}b_\mu^{\alpha}b_\nu^{\beta}A^j_\alpha A^k_\beta)\psi-S^\alpha_{\hp\mu\nu}\partial_\alpha\psi
\end{align}

We try to construct an invariant quantity with minimal modification to \ref{Field strength nabla}; this can be achieved by taking $b_\mu^{\hp\nu}=e^{-f(X)}\delta_\mu^{\hp\nu}$

\begin{equation}
F_{\mu\nu}=\partial_\mu A^i_\nu-\partial_\nu A^i_\mu-S^\sigma_{\hp\mu\nu}A^i_\sigma+ge^{-f}C^i_{\hp jk}A^j_\mu A^k_\nu
\end{equation}

Now we seek out a condition on the torsion to guarantee the invariance of this quantity (more precisely the invariance of $F^i_{\mu\nu}F^{i\mu\nu}$)

\begin{align}
\delta F^i_{\mu\nu}&=-g^{-1}(c_{\nu\hp ,\mu}^{\hp\beta}\theta^i_{,\beta}+c_\nu^{\hp\beta}\theta^i_{,\beta\mu}-c_{\mu\hp ,\nu}^{\hp\beta}\theta^i_{,\beta}-c_\mu^{\hp\beta}\theta^i_{,\beta\nu})+C^i_{\hp jk}(A^k_{\nu ,\mu}\theta^j+A^k_\nu\theta^j_{,\mu}-A^k_{\mu ,\nu}\theta^j-A^k_\mu\theta^j_{,\nu})\nonumber\\
&-S^\beta_{\hp\mu\nu}(-g^{-1}c_\beta^\gamma\theta^i_{,\gamma}+C^i_{\hp jk}A_\beta^k\theta^j)+ge^{-f}C^i_{\hp jk}(-g^{-1}A^j_\mu c_\nu^\gamma\theta^k_{,\gamma}+C^k_{lm}A_\mu^jA^m_\nu\theta^l-g^{-1}c_\mu^\gamma A_\nu^k\theta^j_{,\gamma}\nonumber\\
&+C^j_{lm}A_\mu^m A_\nu^k\theta^l)\\
&=C^i_{\hp jk}\theta^jF^k_{\mu\nu}+g^{-1}(c_{\mu\hp,\nu}^{\hp\lambda}-c_{\nu\hp,\mu}^{\hp\lambda}+S^\alpha_{\hp\mu\nu}c^{\hp\lambda}_\alpha)\theta^i_{,\lambda}
\end{align}

For the coefficient of $\theta^i_{,\lambda}$ to vanish, the torsion must conform with the form;

\begin{equation}
S^\sigma_{\hp\mu\nu}=b_\lambda^{\hp\sigma}(c_{\nu\hp,\mu}^{\hp\lambda}-c_{\mu\hp,\nu}^{\hp\lambda})=\delta_\nu^{\hp\sigma}\chi_{,\mu}-\delta_\mu^{\hp\sigma}\chi_{,\nu}
\end{equation}
which is the form of a vector torsion as defined in \ref{Vtorsion}.

\section{Energy-momentum tensor}\label{AEnergyMomentum}

We need to find the energy momentum tensor for $\Lambda=-\frac{1}{4}F_{\mu\nu}^aF_a^{\mu\nu}$, where

\begin{align*}
\mathscr{T}^\mu_\nu&=q_{,\nu}\frac{\partial\Lambda}{\partial q_{,\mu}}-\delta^\mu_\nu\Lambda\\
&=-F^{\mu\rho}_a F_{\nu\rho}^a+\frac{1}{4}\delta^\mu_\nu F^{\lambda\rho}_a F_{\lambda\rho}^a=-F^{\mu\rho}_a F_{\nu\rho}^a+\delta^\mu_\nu(-\frac{3}{2a^2}(\dot{a}\phi+a\dot{\phi}-a\phi\chi)^2)
\end{align*}

It is now obvious that $\mathscr{T}^\mu_\nu=0$ for $\mu\neq\nu$. For $\mu=\nu$ one explicitly find;

\begin{equation}
\mathscr{T}^\mu_\nu=diag(\rho_A,-P_A,-P_A,-P_A)
\end{equation} 

Where

\begin{align}
\rho_A&\equiv \frac{3}{2}(\dot{\phi}+H\phi-\phi\chi)^2\\
P_A&\equiv \frac{1}{2}(\dot{\phi}+H\phi-\phi\chi)^2
\end{align}
Which is in the form of a perfect fluid namely a radiation field with $P=\frac{1}{3}\rho$.\\
It is due now to point that the usual conservation law of energy-momentum is modified, c.f. \cite{Kranas:2018jdc}. To see that we substitute Eq.(\ref{Friedmann 1}) into Eq.(\ref{Friedmann 2}), and differentiate Eq.(\ref{Friedmann 1});

\begin{equation}\label{dotH sub1}
\dot{H}=-\frac{1}{2}(\rho_A+P_A)+2\chi H-2\dot{\chi}+4\chi^2
\end{equation}

\begin{equation}\label{Freidman 1 dot}
2H\dot{H}=\frac{1}{3}\dot{\rho_A}-4\dot{H}\chi-4H\dot{\chi}-8\chi\dot{\chi}
\end{equation}
multiplying through Eq.(\ref{dotH sub1}) by $2H$ and equating it to Eq.(\ref{Freidman 1 dot});

\begin{align}
\frac{1}{3}\dot{\rho}_A&-4\dot{H}\chi-4H\dot{\chi}-8\chi\dot{\chi}=2H(-\frac{1}{2}(\rho_A+P_A)+2\chi H-2\dot{\chi}+4\chi^2)\nonumber\\
\Rightarrow \dot{\rho}_A&=12\chi(\dot{H}+H^2)+24\chi\dot{\chi}-3H(\rho_A+P_A)+24H\chi^2
\end{align}

Substituting for $\dot{H}+H^2$ from Eq.(\ref{Friedmann 2}), we get the continuity equation for a perfect fluid in the presence of a vector torsion;

\begin{equation}
\dot{\rho}_A=-3H(\rho_A+P_A)-2\chi(\rho_A+3P_A)
\end{equation}

Which reduces to the usual continuity equation in the limit of a vanishing torsion.

\section{Ricci and Einstein tensors}\label{ARIcci and E tensors}

Explicit calculation for the Ricci tensor produces the following non-vanishing components

\begin{align}
R_{00}&=\nonumber -3\big[\frac{\ddot{a}}{a}+2\dot{\chi}+3\chi\frac{\dot{a}}{a}\big]\\
R_{11}&=\nonumber\frac{1}{1-Kr}(a\ddot{a}+2\dot{\chi}a^2+2\dot{a}^2+10\chi a\dot{a}+8\dot{\chi}a^2+2K)\\
R_{22}&=\nonumber r^2(a\ddot{a}+2\dot{\chi}a^2+2\dot{a}^2+10\chi a\dot{a}+8\dot{\chi}a^2+2K)\\
R_{33}&=\nonumber r^2\sin^2(\theta)(a\ddot{a}+2\dot{\chi}a^2+2\dot{a}^2+10\chi a\dot{a}+8\dot{\chi}a^2+2K)
\end{align}
And the Ricci scalar;

\begin{equation}
R=\frac{6}{a^2}(a\ddot{a}+\dot{a}^2+6\chi a\dot{a}+2a^2(2\chi^2+\dot{\chi})+K)
\end{equation}

The nonvanishing components of Einstein tensors are;

\begin{align}
G_{00}&=\nonumber 3\big[\big(\frac{\dot{a}}{a}\big)^2+4\chi\frac{a\dot{a}}{a^2}+4\chi^2+\frac{K}{a^2}\big]\\
G_{11}&=\nonumber \frac{-1}{1-Kr^2}(2a\ddot{a}+8\chi a\dot{a}+\dot{a}^2+4\chi^2a^2+4\dot{\chi}a^2+K)\\
G_{22}&=\nonumber -r^2(2a\ddot{a}+8\chi a\dot{a}+\dot{a}^2+4\chi^2a^2+4\dot{\chi}a^2+K)\\
G_{11}&=\nonumber -r^2\sin^2(\theta)(2a\ddot{a}+8\chi a\dot{a}+\dot{a}^2+4\chi^2a^2+4\dot{\chi}a^2+K)
\end{align}

\clearpage
\bibliographystyle{unsrt}
\bibliography{Biblography}

\end{document}